\documentstyle[aps,graphicx,floats]{revtex} 
\def\GeV{\mbox{ GeV}} 
\def\TeV{\mbox{ TeV}} 
\def\fb{\mbox{ fb}} 
 
\begin{document} 
\def\beq{\begin{eqnarray}}    
\def\eeq{\end{eqnarray}}      
\def\mg{m_{\tilde{g}}} 
\def\msb1{m_{\tilde{b}_{1}}} 
\def\mst1{m_{\tilde{t}_{1}}} 
\draft 
\twocolumn[\hsize\textwidth\columnwidth\hsize\csname 
@twocolumnfalse\endcsname 
\title{Prospects for supersymmetric charged Higgs boson discovery\\ at the Tevatron  and the LHC} 
 
\author{Alexander Belyaev\( ^{a} \), David Garcia\( ^{b} \), Jaume Guasch\( ^{c} \), 
Joan Sol\`{a}\( ^{d} \)} 
\address{\textit{\( ^{a} \)\, 
Physics Department,Florida State University, 
Tallahassee, FL 32306-4350, USA}\\ 
\textit{\( ^{b} \)\,Theory Division, CERN, CH--1211 Geneva 23, 
Switzerland}\\ \textit{\( ^{c} \)\,Institut f\"{u}r Theoretische 
Physik, Universit\"{a}t Karlsruhe, Kaiserstra\ss{}e 12, D--76128 
Karlsruhe, Germany} 
\\ \textit{\( ^{d} \)\, Grup de F\'{\i}sica Te\`{o}rica and Institut de F\'{\i 
}sica d'Altes Energies,}\\ 
\textit{Universitat Aut\`{o}%
noma de Barcelona, E--08193, Bellaterra, Barcelona,Catalonia, 
Spain} } 
 
\maketitle 
 
\begin{abstract} 
We investigate the prospects for heavy charged Higgs boson 
production through the mechanisms $p\bar{p}(pp)\to\,tbH^{\pm} +X$ 
at the upgraded Fermilab Tevatron and at the upcoming LHC collider 
at CERN respectively. We focus on the MSSM case at high values of 
$\tan \beta\gtrsim m_{t}/m_{b} $ and include the leading SUSY 
quantum corrections. A detailed study is performed for all 
important production modes and basic background processes for the 
\( t\bar{t}b\bar{b} \) signature. At the upgraded Tevatron a 
charged Higgs signal is potentially viable in the $220-250\GeV$ 
range or excluded at 95\,\%CL up to $300\GeV$. At the LHC, a 
$H^{\pm}$ of mass up to $800\GeV$ can be discovered at $5\sigma$ 
or else be excluded up to a mass of $\sim 1.5\TeV$. The presence 
of SUSY quantum effects may highly influence the discovery 
potential in both machines and can typically shift these limits by 
$200\GeV$ at the LHC. 
\end{abstract} 
 
\pacs{PACS: 12.60.Fr, 13.85.-t \hspace*{0.3cm}
 FSU-HEP-050101, KA-TP-15-2001, UAB-FT-512 
 \hspace*{0.3cm} hep-ph/0105053}
 ]
 
 
The full experimental confirmation of the Standard Model (SM) is 
still waiting for the finding of the Higgs boson. Last LEP 
results, suggesting a light neutral Higgs of about 
$115\GeV$~\cite{LEP115}, are encouraging, but we will have to wait 
the news from the upgraded Fermilab Tevatron or from the upcoming 
Large Hadron Collider at CERN to see this result either confirmed 
or dismissed. For intermediate masses above the LEP limit and 
below $180\GeV$ there is a chance for the Tevatron, but for 
higher masses up to $1\TeV$ one needs the LHC. However, even if a 
neutral Higgs boson is discovered, the principal question will 
stand immutable at the forefront of Elementary Particle research: 
is the minimal SM realized in nature or does a model beyond the SM 
exist with an extended Higgs sector? In most of these extensions, 
the physical spectrum contains neutral Higgs particles and some of 
them may mimic the SM one.  But in general they also involve 
charged Higgs bosons, and this introduces an obvious distinctive 
feature. For example, in the general two-Higgs-doublet model 
(2HDM)~\cite{Hunter} one just adds up another doublet of Higgs 
scalars and then the spectrum of the model contains three neutral 
Higgs bosons $h=h^0,H^0,A^0$ and two charged ones $H^{\pm}$. While 
the detection of a charged Higgs boson would still leave a lot of 
questions unanswered, it would immediately offer (in contrast to 
the detection of a neutral one) indisputable evidence of physics 
beyond the SM. In this Letter we report on the main results from a 
fully-fledged study of the $H^{\pm}$ production in hadron 
colliders. We restrict our analysis to the Higgs sector of the 
Minimal Supersymmetric Standard Model (MSSM), which is beyond 
doubt the most prominent (Type II) 2HDM~\cite{Hunter}. The lengthy 
details of this work will be presented elsewhere. 
 
The relevant mechanisms on which we will concentrate 
 \beq 
 p\bar{p}(pp)\to\,tbH^{\pm} +X\label{TeV} \ \ ({\rm Tevatron})({\rm LHC})\,, 
\label{signal} 
 \eeq 
are long known to be the leading ones for $H^{\pm}$ production at 
high $\tan\beta$\,\cite{Gunion}. They constitute the charged 
counterpart of the process $p\bar{p}\to\,t\bar{t}H +X$ for Higgs 
boson production in the SM, recently revisited in \cite{Goldstein} 
for the Tevatron, and of  $p\bar{p}(pp)\to\,t\bar{t}h +X$ for 
neutral MSSM Higgs boson production at the Tevatron and the LHC 
\cite{Mrenna}. While the study of (\ref{signal}) has been further 
addressed in the literature\,\cite{BorzuMiller}, to the best of 
our knowledge all the works on this subject-- except a first 
estimation in \,\cite{THWG}-- do stick to a tree-level computation 
without SUSY quantum effects, in spite of the fact that some of 
them explicitly admit that the sort of charged Higgs boson they 
are dealing with is of the MSSM type. Therefore, they are 
unavoidably affected by some drawbacks. In the present work we 
have implemented several additional features which improve in a 
substantial manner our knowledge on the real capability for the 
mechanisms (\ref{signal}) to produce a charged Higgs boson or to 
put limits on its mass within the MSSM. First, we  include the 
leading SUSY radiative corrections (both strong and electroweak) 
along with an analysis of the off-shell effects. Second, we 
perform a beyond-the-parton-level simulation of events, which 
includes the toy-detector simulation, jet fragmentation, and 
initial and final radiation effects. And yet another new aspect of 
the present work is the proper kinematical analysis of the 
$gg\rightarrow H^{+}\bar{t}b$ and $g\bar{b}\rightarrow 
H^{+}\bar{t}$ subprocesses at the level of differential 
cross-sections after correctly combining them, using the standard 
recipe for the total cross-section\,\cite{Dicus}, and a method 
based on Ref.~\cite{BelaBoos} for the differential one. 
 
A realistic study of charged Higgs boson physics cannot be 
accomplished without including the information provided  by 
radiative corrections. These are not only potentially large in the 
computation of the MSSM Higgs boson masses themselves but also in 
the interaction vertices and self-energies of given processes, 
particularly in the decay of the top quark into charged Higgs and 
for the hadronic decays of the MSSM Higgs 
bosons~\cite{SUSYtbH,eff}. 
\begin{figure}[tb] 
\centerline{\resizebox{!}{4.8cm}{\includegraphics{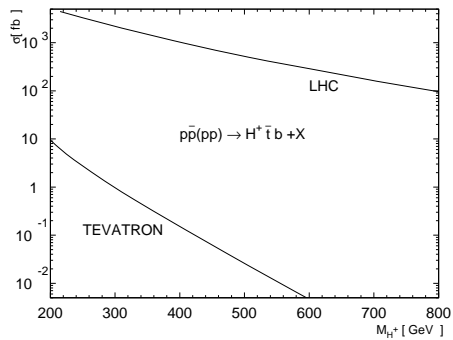}} }
\caption{Tree-level cross-sections (\ref{signal}) versus 
$M_{H^{\pm}}$ at the Tevatron Run II ($2\TeV)$ and LHC 
($14\TeV)$ for $\tan\beta=50$.\label{fig_cs}} 
\end{figure} 
\begin{figure}[tb] 
\centerline{\resizebox{!}{4.8cm}{\includegraphics{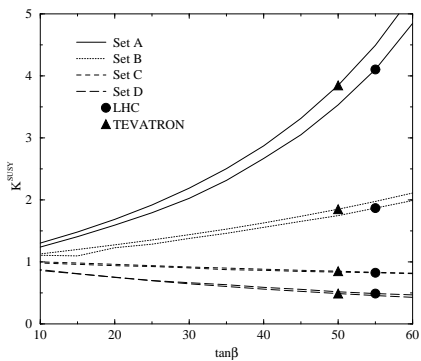}}}
\caption{$K_{SUSY}=\sigma^{\rm SUSY}/\sigma^{\rm tree}$ factor for 
the signal (\ref{signal}) as a function of $\tan\beta$ for the 
typical sets of parameters in Table\,\protect\ref{Input}, at the 
Tevatron (for $M_{H^{\pm}}=250\GeV$) and the LHC (for 
$M_{H^{\pm}}=500\GeV$).} \label{fig1} 
\end{figure} 
\begin{table}[tb] 
\begin{tabular}{|c|c|c|c|c|c|c|c|} 
& $\mu$ & $M$ & $\mg$ &$\mst1$ & $\msb1$ & $A_t$ & $A_b$ 
\\\hline 
Set A &  -1000 & 200 & 1000 & 1000 & 1000 &  500 & 500\\\hline 
Set B &   -200 & 200 & 1000 &  500 &  500 &  500 & 500\\\hline 
Set C &    200 & 200 & 1000 &  500 &  500 & -500 & 500\\\hline 
Set D &   1000 & 200 & 1000 & 1000 & 1000 & -500 & 500 
\end{tabular} 
\caption{Typical sets of SUSY parameters used in the computation 
of the signal cross-section (\ref{signal}) in Fig.\,\ref{fig1} 
(all entries in GeV). Here $\mu$ and $M$ are the higgsino and 
$SU(2)$ gaugino mass parameters, $\mg$ is the gluino mass, $\mst1$ 
and $\msb1$ are the lightest stop and sbottom masses, and $A_t$, 
$A_b$ are the top and bottom quark trilinear  SUSY-breaking 
couplings. The $|\delta\rho|<0.001$ constraint is satisfied. 
Notation as in Ref.\protect\cite{SUSYtbH}.} \label{Input} 
\end{table} 
\noindent Supersymmetric quantum effects can also be very 
important for the processes (\ref{signal}). Indeed, in 
Fig.~\ref{fig_cs} we present the tree-level signal cross-sections 
as a function of $M_{H^{\pm}}$ for the LHC and Tevatron colliders, 
whereas in Fig.\,\ref{fig1} a rich variety of SUSY effects is 
exhibited, as a function of $\tan\beta$, for the various sets of 
MSSM parameters indicated in Table\,\ref{Input}. For the signal 
the leading corrections can be described through an effective 
Lagrangian approach\,\cite{eff} 
\begin{equation} 
{\cal L}=\frac{gV_{tb}}{\sqrt{2\,}M_{W}}\,\frac{\overline 
{m}_{b}\,\tan\beta}{1+\Delta m_{b}}\,H^{+}\overline{t}_{L}%
\,b_{R}+h.c. \label{effecLag}%
\end{equation} 
where $\overline{m}_{b}$ is the running bottom mass in the 
$\overline{MS}$. The previous formula allows to treat the leading 
SUSY Yukawa coupling effects correctly resummed to all 
orders\,\cite{eff}. The analytic form of the strong (SUSY-QCD) and 
electroweak (SUSY-EW) corrections $\Delta m_{b}$ in the MSSM is 
given in Ref.\cite{eff}. Although $\Delta m_{b}$ is the only 
correction that contributes at order $(\alpha/4\pi)\,\tan\beta$ 
($\alpha=\alpha_{S},\alpha_{W}$) and thus dominates for large 
$\tan\beta$, we have also included off-shell SUSY-QCD and SUSY-EW 
corrections to the $tbH^{\pm}$ vertex and to the fermion 
propagators. We have made extensive use of the CompHEP package for 
the algebraic and numerical calculations\,\cite{CompHEP}. Despite 
CompHEP is only able (in principle) to deal with tree-level 
calculations, with the help of Eq.\,(\ref{effecLag}) we have 
managed to add the SUSY corrections to the $tbH^{\pm}$ vertex and 
fermion propagators and we have assessed the relevance of the 
off-shell contributions. To this end we have evaluated the full 
set of one-loop SUSY diagrams for the relevant $tbH^{+}$ vertex. 
The same set was considered in detail in Ref.\cite{SUSYtbH} for 
the case where all external particles are on-shell. In the present 
instance, however, at least one of the quarks in that vertex is 
off shell. Therefore, we can use the same bunch of diagrams as in 
the on-shell case but we have to account for the off-shell 
external lines, which is a non-trivial task. We have studied this 
issue in detail by expanding the off-shell propagators. First, we 
have modified CompHEP's Feynman rules to allow for the most 
general off-shell $\bar{t}bH^{+}$ vertex; then we have let CompHEP 
reckon the squared matrix elements and dump the result into REDUCE 
code. The subsequent numerical computation of the one-loop 
integrals has been done using the package 
LoopTools~\cite{LoopTools}. At this point, we have inserted 
expressions for the coefficients of the off-shell $\bar{t}bH^{+}$ 
vertex that include the one-loop off-shell supersymmetric 
corrections to the vertex itself and to the off-shell fermion 
propagators and fermionic external lines. Only half the 
renormalization of an internal fermion line has to be included, 
the other half being associated to the $gqq$ vertex. This 
procedure has allowed us to estimate the relative size of the 
off-shell effects in the signal cross-section, which never exceeds 
the few per cent level. The upshot is that the approximation of 
neglecting vertex and propagator corrections in the cross-section, 
which may be called ``improved Born'' approximation, is really 
justified in the relevant region of parameter space. 
 
 
\begin{figure}[tb] 
\vskip -0.5cm \centerline{\resizebox*{6cm}{!}{\includegraphics{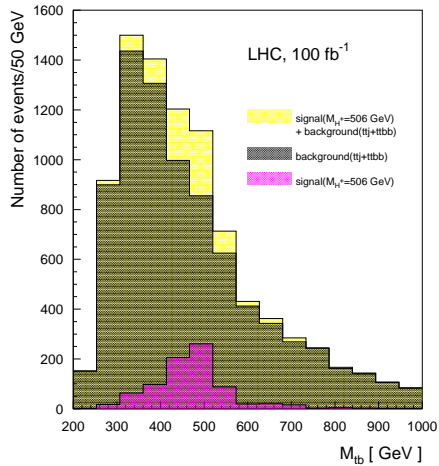} } }
\caption{Reconstructed $tb$ invariant-mass distribution for 
signal, background and signal plus background events, for 
$M_{H^{\pm}}=500\GeV$, at the LHC.\label{fig:tb}} 
\end{figure} 
As for the signal versus background study we have focused on the 
$t\bar{t}b\bar{b}$ signature corresponding to the 
$H^{+}\rightarrow t\bar{b}$ decay channel, which has the biggest 
branching ratio at high $\tan\beta$. Then we have concentrated on 
the triple-$b$ tagging case, which gives the best possibility to 
measure the signal cross-section (\ref{signal}). This is a crucial 
point, especially for the Tevatron, where the production rate is 
too small to give any viable signal in the case of four-$b$ 
tagging. As regards the LHC, a triple-$b$ tagging study allows the 
signal cross-section to be measured more precisely, even though 
the signal/background ratio can be better for the four-$b$ tagging 
case. We notice that the background processes in Table\,\ref{bkg} 
are insensitive to the leading type of SUSY effects that 
contribute to (\ref{signal}). To perform a realistic signal and 
background event simulation we complied with the following 
procedure. The (tree-level) matrix elements for the signal and 
background processes have been calculated using the CompHEP 
package~\cite{CompHEP}. The next step was the parton-level event 
simulation, also with the help of CompHEP. Then we automatically 
linked the parton-level events from CompHEP to the PYTHIA 6.1 
Monte Carlo generator and the CompHEP--PYTHIA 
interface~\cite{pythia}. Therefore we took into account the 
effects of the final-state radiation, hadronization and string-jet 
fragmentation using PYTHIA tools. The following resolutions were 
used for the jet and electron energy smearing: $\Delta 
E^{had}/E=0.8/\sqrt{E} $ and $\Delta E^{ele}/E=0.2/\sqrt{E} $. In 
our analysis we used the cone algorithm for the jet reconstruction 
with a cone size $\Delta 
R=\sqrt{\Delta\varphi^{2}+\Delta\eta^{2}}=0.7 $. The choice of 
this jet-cone value is related to the crucial role of the 
final-state radiation (FSR), which strongly smears the shape of 
the reconstructed charged Higgs boson mass. We have checked that 
the value of 0.7 minimizes the FSR effects. Now, in order to 
decide whether a charged Higgs cross-section leads to a detectable 
signal, we have to compute the background rate. 
Since the  mistagging probability of light quark and gluon jets 
is expected  to be $\lesssim 1\% $  \cite{cmstag}, the only 
significant backgrounds leading to the same $t\bar{t}
$ signature are those  shown in Table~\ref{T2} along with their 
respective cross-sections. For the $t\bar{t}b\bar{b}$ and $t\bar t 
qg$ processes we have applied the jet  separation cut 
$\Delta_{R}^{jj}>0.5(\Delta_{R}=\sqrt{\Delta\theta^{2}+\Delta\phi^{2}})$ 
and the initial  cut $p_{T}^{j}>10\GeV$ ($p_{T}^{j}>20\GeV$) at 
the Tevatron (LHC). For the $t\bar{t}j$ process  the initial cut 
$p_{T}^{j}>10\GeV$ ($p_{T}^{j}>20\GeV$) was applied at  the 
Tevatron (LHC). To obtain a realistic description of the $b 
$-tagging efficiency as a function of $b$-quark transverse 
momenta, for the Tevatron we use the projected $b$-tagging 
efficiency of the upgraded D\O\ detector\,\cite{FNAL98} while for 
the LHC we parameterize numerical results from the CMS 
collaboration\cite{cmstag}. Efficiencies for both 
parameterizations are about $60\%$ at the $p_T^b$ saturation value 
of $\sim 100\GeV$. 
We assume that $b $-jets can be tagged only for pseudorapidity $|\eta_{b}%
|\leq2 $ by both Tevatron and LHC experiments. Furthermore, we 
have optimized the reconstruction procedure, the $p_T$ cut on the 
leading $b$-jet ($p_{T}^{b}>[M_{H^{+}}/5-15]\mbox{GeV}$) and 
window cut on the $tb$-invariant mass around the selected values 
of $M_{H^{+}}$ ($|m_{tb}-M_{H^{+}}|<5\sqrt{M_{H^{+}}}$) to achieve 
the maximal significance of the cross-section signal $\sigma_S$ 
(\ref{signal}) versus the background, $\sigma_S/\sqrt{B}$. The 
typical efficiency at the Tevatron is $5-6\%$ while for LHC it 
goes down to $1-2\%$. These values include the triple $b$-tagging, 
branchings of $W$-bosons decays (leptonic and hadronic decay 
modes) and the efficiency of the kinematical cuts and 
reconstruction of the $t\bar{t}b\bar{b}$ signature. As an example, 
Figure~\ref{fig:tb} shows the reconstructed $tb$ invariant-mass 
distribution for signal and background events at the LHC. 
 
\begin{figure}[tb] 
\centerline{\resizebox{!}{4cm}{\includegraphics{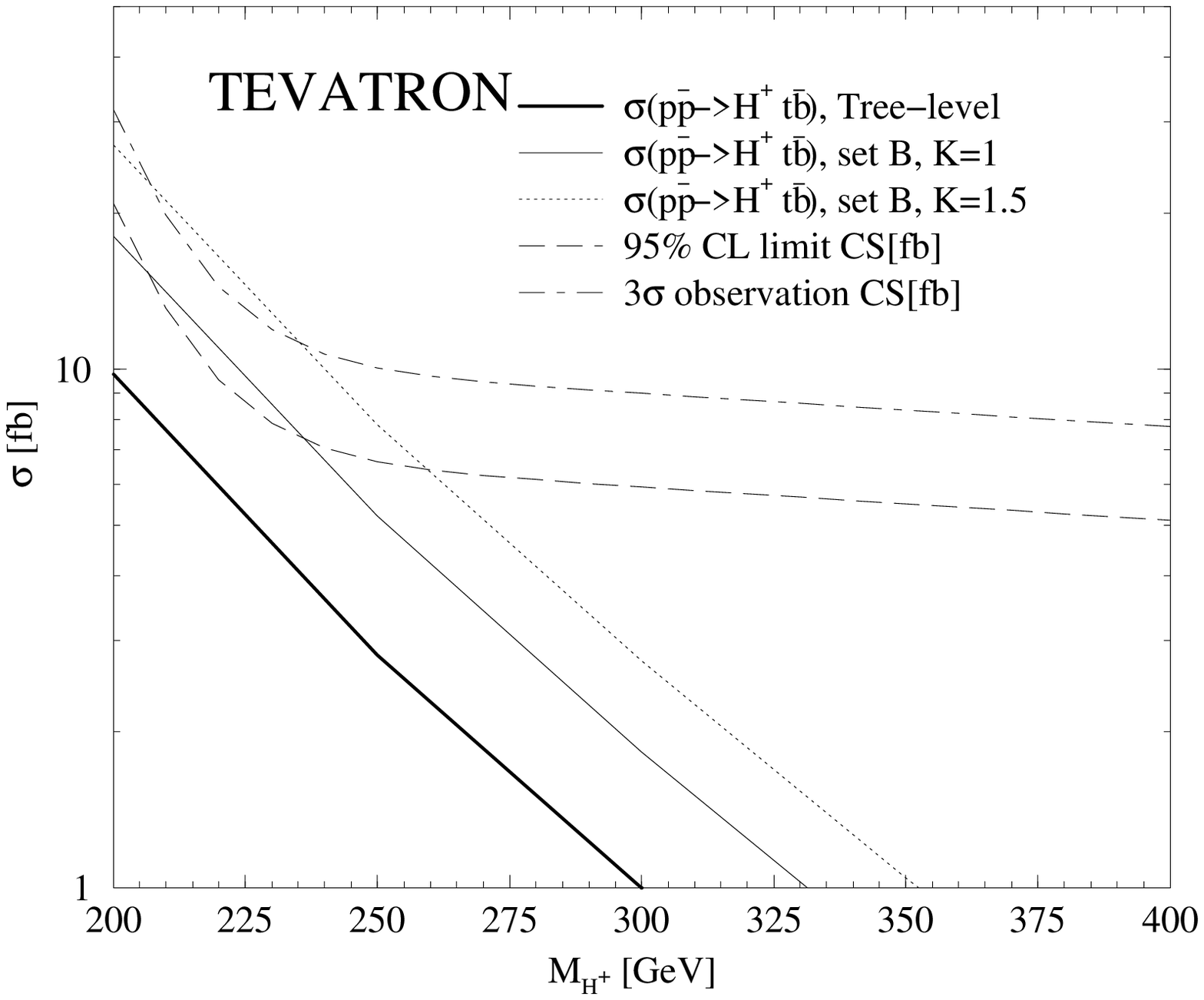}}}
\centerline{\resizebox{!}{4cm}{\includegraphics{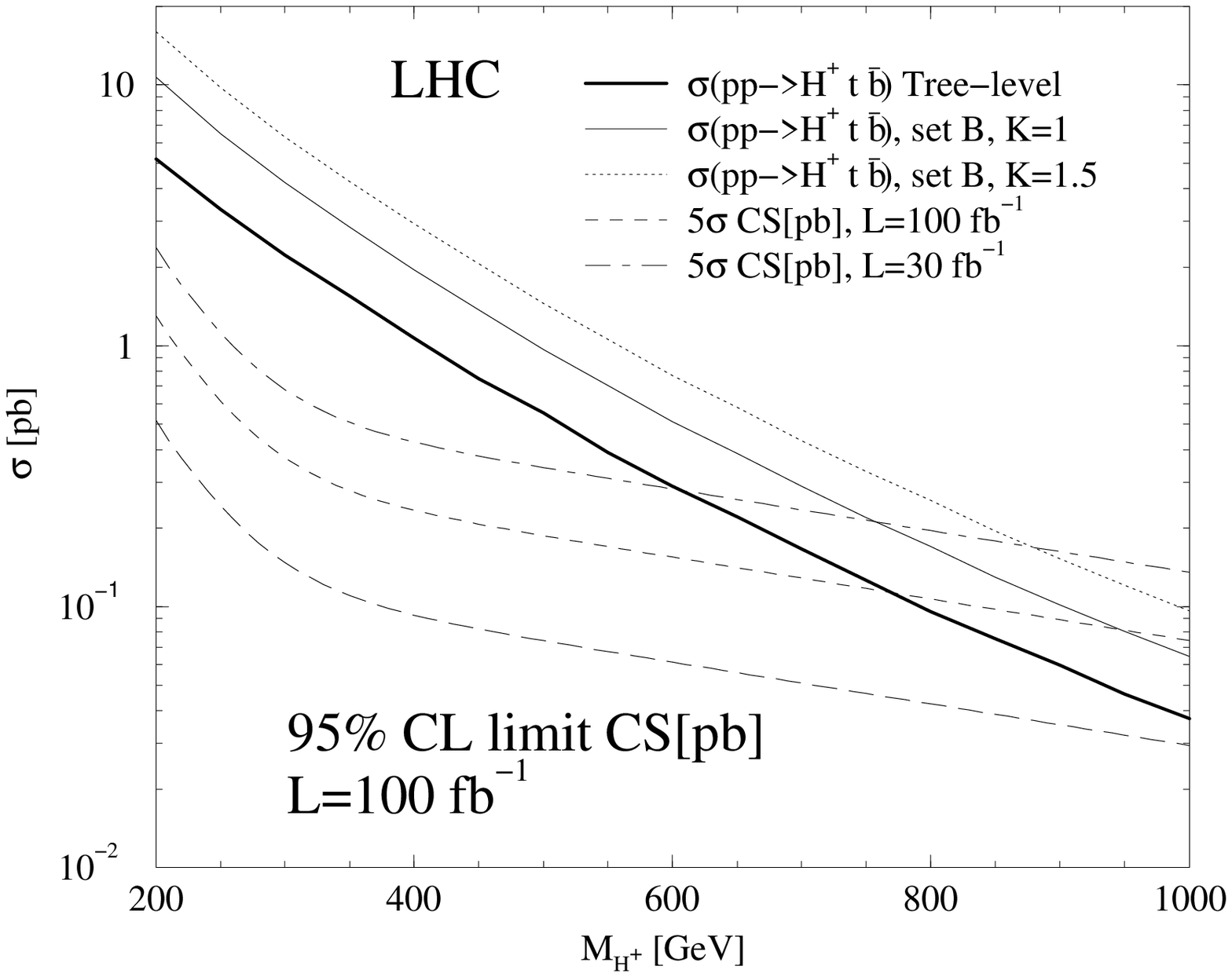}}}
\caption{Discovery and exclusion limits for the 
charged Higgs boson at the (a) Tevatron and (b) LHC,  for 
$\tan\beta=50$. Shown are the total cross-sections for: i) the 
signal (\ref{signal}) at the tree-level, ii) the SUSY corrected 
signal for the input set B in Table\,\protect\ref{Input} including 
QCD factors $K=1$ and $K=1.5$, iii) the $3\,\sigma$ (Tevatron) or 
$5\,\sigma$ (LHC) discovery limits at the integrated luminosities 
$L=\int{\cal L} dt=30~fb^{-1}$ and $100~fb^{-1}$, iv) the 95\% CL 
exclusion limit at $L=100~fb^{-1}$.} \label{fig2} 
\end{figure}

In Fig.\,\ref{fig2} we present the discovery and exclusion limits 
for $H^{\pm}$ at the Tevatron and the LHC at high $\tan\beta$: It 
shows the signal cross-section $\sigma_S$ and the cross-sections 
which would lead to the $5\sigma$, $3\sigma$ and $1.98\sigma$(95\% 
CL) significance. From the intersections of the last three with 
$\sigma_S$ we infer the mass ranges which can be discovered or 
excluded. The tree-level case is also shown, and it is seen to be 
too small at the Tevatron to place any limit. But when including 
SUSY effects the situation changes. We take the moderate input set 
B from Table\,\ref{Input}. The unknown QCD effects at NLO are 
estimated by including a $K$-factor (typically one expects 
$K\simeq 1.5$\,\cite{BorzuMiller}), and we used a bottom quark 
pole mass $m_b=4.6\GeV$. 
\begin{table} 
\vskip -0.3cm 
\begin{tabular}{llll} 
(a) & $\sigma(qq\rightarrow t\bar{t}b\bar{b}) $  &  6.62 fb & 0.266 pb \\ 
& $\sigma(gg\rightarrow t\bar{t}b\bar{b}) $  &  0.676 fb &   6.00 pb \\ 
& $\sigma(gb\rightarrow t\bar{t}b) $  &  1.22 fb  & 4.33 pb \\ 
& Subtr. term  &  0.72 fb &   2.1 pb\\ 
 \\ 
(b) & $\sigma(q\bar{q}\rightarrow g\bar{t}\bar{t}) $  &  1890 fb & 21 pb \\ 
& $\sigma(gq\rightarrow qt\bar{t})$  & 193 fb &   122 pb \\ 
& $\sigma(gg\rightarrow g\bar{t}\bar{t}) $ &  262 fb  & 371 pb \\ 
\end{tabular} 
\vspace{0.15in} \caption{(a)\, The main background processes to 
the signal (\ref{signal})  at the Tevatron (2nd column) 
and the LHC (3rd column) under the cuts explained in the text. The 
various contributions are shown together with that of the 
subtraction term\,\protect\cite{Dicus}; (b)\, Background from 
$pp\rightarrow t\bar{t}q{g}$ when the light quark or gluon are 
misidentified as a $b$-jet. } \vspace{0.15in} \label{T2} 
\label{bkg} 
\end{table} 
\noindent For example, for $M_{H^{\pm}}=215\GeV$ at the Tevatron 
one would expect $7$ signal and about $6$ background reconstructed 
events at $L=25\fb^{-1}$. At the LHC with $M_{H^{\pm}}=500\GeV$ we 
have $1200$ and $3800$ signal and background events respectively 
at $L=100\fb^{-1}$. A canonical $5\,\sigma$ discovery limit around 
$800\GeV$ can be obtained at the LHC for the MSSM charged Higgs, 
or else an exclusion limit at $95\%\,C.L.$ up to at least 
$1.2\TeV$ ($K=1$). For the Tevatron we can now place a 
$95\%\,C.L.$ exclusion limit in the mass range $200-250\GeV$, if 
$K=1$. However, if the QCD factor is $K\simeq 1.5$ this would open 
the exciting opportunity to observe the charged Higgs already at 
the Tevatron with a $3\sigma$ significance in the mass range 
$220-250\GeV$!. Moreover, one should notice that $b$-tagging 
efficiency could be increased up to $\sim 70\%$ with the use of 
the $3$D vertexing algorithms ~\cite{Goldstein}. For triple 
$b$-tagging, this would augment the signal by at least a factor of 
$2$, and so the discovery and the $95\%$ exclusion limits would be 
extended accordingly. Needless to say, in case of maximal SUSY 
enhancement (Cf. set A) the exclusion/observation mass range for 
the charged Higgs boson would be further enlarged. 
 
In summary, we have assessed the possibility to see a SUSY charged 
Higgs at the Tevatron and the LHC  through the mechanisms 
(\ref{signal}). Our study of the quantum corrections within the 
MSSM has shown that at high $\tan\beta$ they are dominated by 
exceptionally important effects that can be absorbed into an 
effective $tbH^{\pm}$ vertex, and therefore in practice they can 
be treated as an ``improved Born approximation''. From a detailed 
signal versus background study we have shown that the prospects 
for the signal discovery are viable for the Tevatron (if 
$M_{H^{\pm}}\lesssim 250\GeV$) and promising for the LHC (if 
$M_{H^{\pm}}\lesssim 1.5\TeV$). In deriving these limits we have 
demonstrated that the quantum machinery from the MSSM can play a 
crucial role to increase the signal. In large portions of the 
parameter space the genuine SUSY corrections could show up as a 
smoking gun over the tree-level cross-section (after subtracting 
the conventional QCD effects) with a contribution of the order or 
larger than the QCD effects themselves. Since both the processes 
(\ref{signal}) and the SUSY effects are only relevant at high 
$\tan\beta$, the sole presence of the signal could be a hallmark 
of the underlying MSSM dynamics. A close comparison with the 
neutral Higgs boson production processes mentioned above should 
be, if available, very useful to confirm or dismiss the MSSM 
nature of the charged and neutral Higgs bosons. 
 
\noindent 
The work of D.G. and J.G. has been supported by the E.U. Marie 
Curie program, and that of J.S. by CICYT. A.B. thanks J. Owens,
N. Kidonakis and L. Reina for stimulating discussions and acknowledges the 
support of U.S. DOE. \vskip -0.5cm 
 
\end{document}